\documentclass{ws-procs9x6-cpt22}
\begin{document}

\newcommand{\refeq}[1]{(\ref{#1})}
\def\etal {{\it et al.}}

\title{Lorentz Tests with Astrophysical Photons}

\author{F.\ Kislat}

\address{Department of Physics \& Astronomy and Space Science Center,\\
University of New Hampshire,\\
Durham, NH 03824, USA}

\begin{abstract}
Astrophysical polarization measurements provide some of the strongest constraints on the photon sector of the non-minimal Standard-Model Extension.
This paper reviews some recent results obtained by combining optical linear and circular polarization data from a large number of astrophysical objects.
\end{abstract}

\bodymatter

\section{Introduction}
When photons emitted by astrophysical objects travel over a long distance before being observed at Earth, minute deviations of their phase velocity from $c$ accumulate, potentially resulting in observable effects.
This makes astrophysical observations some of the most sensitive probes of Lorentz Invariance Violation. 
In the photon sector of the non-minimal Standard-Model Extension (SME), Lorentz and CPT violating effects are quantified by a set of coefficients, which for astrophysical tests are most usefully represented by an expansion in mass-dimension $d$ and spherical harmonics.\cite{sme}
This representation allows a classification of models into birefringent CPT-odd models (odd $d$, SME coefficients $k_{(V)jm}^{(d)}$), birefringent CPT-even models (even $d$, coefficients $k_{(E)jm}^{(d)}$ and $k_{(B)jm}^{(d)}$), and non-birefringent CPT-even models (coefficients $c_{(I)jm}^{(d)}$).
Various considerations constrain the values of $j$ allowed for a given set of coefficients and mass-dimension $d$.
Furthermore, a massless photon requires that all coefficients $k_{(E)jm}^{(d)}$ and $k_{(B)jm}^{(d)}$ with $j<2$ vanish.
This implies that any Lorentz-violating, birefringent, CPT-even model must result in anisotropies of the phase velocity of light.

\begin{figure}[t]
    \centering%
    \includegraphics[width=.8\linewidth]{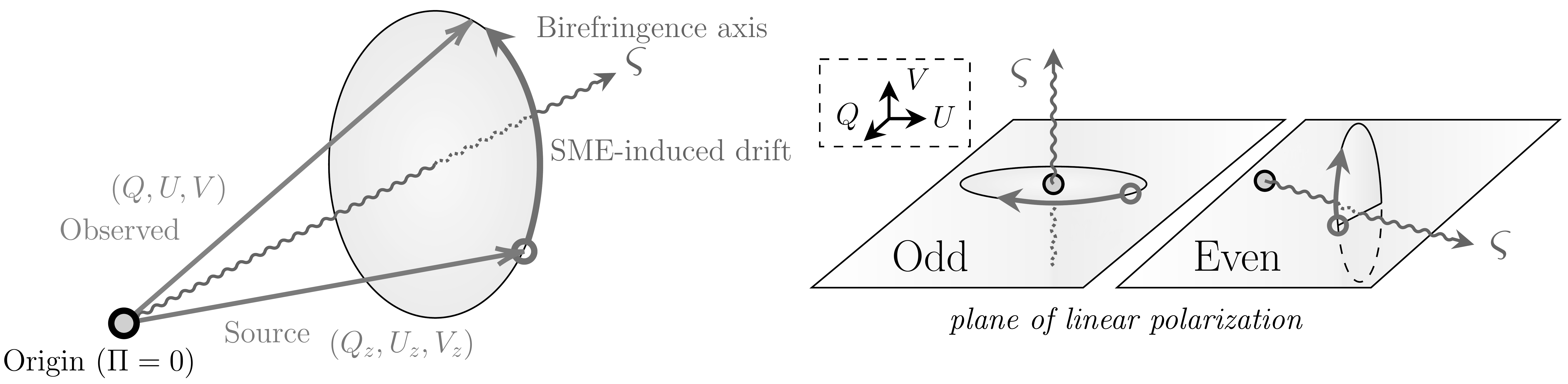}%
    \caption{\emph{Left:} The Stokes vector describes a cone around the birefringence axis $\boldsymbol{\varsigma}$.
    \emph{Right:} In odd-$d$ models, $\boldsymbol{\varsigma}$ is aligned with the Stokes $V$-axis; in even-$d$ models, $\boldsymbol{\varsigma}$ lies in the $Q$-$U$ plane.
    Adapted from Friedman \etal\ (2020).\cite{friedman_etal_2020}}
    \label{fig:cones}
\end{figure}

Measurements of the polarization of light from distant astrophysical objects provide a particularly powerful test.
Such measurements are sensitive to the phase difference between different polarization modes, not limited by the time resolution achievable by time-of-flight experiments.
As a photon propagates, its Stokes vector $\mathbf{s} = (Q,U,V)^T$ varies,
\begin{equation}\label{eq:dsdt}
    \tfrac{d\mathbf{s}}{dt} = 2E\boldsymbol{\varsigma}\times\mathbf{s},
\end{equation}
where $E$ is the photon energy and $\boldsymbol{\varsigma} = (\varsigma^1,\varsigma^2,\varsigma^3)^T$ is a vector in Stokes space determined by the SME coefficients and depending on photon energy as $E^{d-4}$ as well as its direction of propagation.
Over time, the Stokes vector $\mathbf{s}$ describes a cone around the birefringence axis $\boldsymbol{\varsigma}$ as illustrated in Fig.~\ref{fig:cones} (\emph{left}).
As shown in Fig.~\ref{fig:cones} (\emph{right}), the birefringence axis $\boldsymbol\varsigma$ is aligned with the Stokes $V$-axis in the odd-$d$ case, whereas in the even-$d$ case it lies in the $Q$-$U$ plane of linear polarization.

Due to the energy dependence in Eq.~\refeq{eq:dsdt} and the energy dependence of~$\boldsymbol\varsigma$, the rate of change of the Stokes vector depends on photon energy.
Thus, a signature of Lorentz violation is an energy-dependence of the observable polarization.
Additionally, due to this energy dependence, observations at high frequencies have the greatest sensitivity to Lorentz violating effects.
In spectropolarimetric measurements where energy-resolved polarization information is available, such a variation of the Stokes parameters is, in principle, directly measurable.
However, often only spectrally integrated polarization data is available.
Furthermore, beyond radio frequencies, only few measurements of circular polarization exist.

\section{Optical Polarization of AGN and GRB afterglows}
In this section we summarize constraints on SME coefficients we have obtained from optical polarization measurements of distance astrophysical objects.
While a direct detection of LIV signatures is in principle possible, matters are complicated by uncertainties about the emission properties at the source.
Therefore, all our analyses have been constructed to only provide limits on SME coefficients.

In case of odd $d$, the orientation of $\boldsymbol\varsigma$ along the $V$ axis implies that circular polarization is unaffected during propagation, whereas linear polarization will continuously rotate resulting in a constantly changing polarization direction.
In case of spectrally integrated measurements, this effect will reduce the polarization degree from its value at the source as the measurement will be integrating over different polarization directions.
In fact, for sufficiently large SME coefficients, no linear polarization would be observable.
We have used the absence of changing polarization angle in spectropolarimetric measurements and detection of non-zero optical polarization from distant AGN and gamma-ray burst afterglows to place limits on $d=5$ SME coefficients.\cite{friedman_etal_2020,kislat_krawczynski_2017}

The even-$d$ case is more complicated due to the birefringence axis in the plane of linear polarization.
As Fig.~\ref{fig:cones} (\emph{right}) illustrates, even a purely linearly polarized wave at the source will gradually acquire some circular polarization $V$ while maintaining a constant total polarization fraction $\Pi = \sqrt{Q^2+U^2+V^2}$.
At the same time, the linear polarization angle will oscillate between two values determined by the extremes of the $Q$ and $U$ components of $\mathbf{s}$.
The degree to which this transformation between linear and circular occurs, and the amplitude of linear polarization changes is determined by the angle between $\mathbf{s}$ and $\boldsymbol\varsigma$ at the source.
In addition to the values of SME coefficients, the effect thus depends on the direction to the source, photon energy, and polarization properties at the source.
The latter are generally unknown.

In order to constrain birefringent $d=4$ and $d=6$ SME coefficients, we collected linear polarization data from a large sample of sources.
For a given point in the $\left[2(d-1)^2-8\right]$-dimensional SME coefficient space we calculate the likelihood to make each of the measurements.
This is possible even though the true source polarization angle is unknown, because this angle is directly related to the observed polarization angle for a given set of SME coefficients.
We use the combined likelihood of all observations to sample the coefficient space using a Markov Chain Monte Carlo (MCMC) technique.
This allows us to calculate the posterior probability distribution for each of the coefficients.\cite{friedman_etal_2020, kislat_2018}

\begin{figure}[t]
    \centering%
    \includegraphics[width=.75\linewidth]{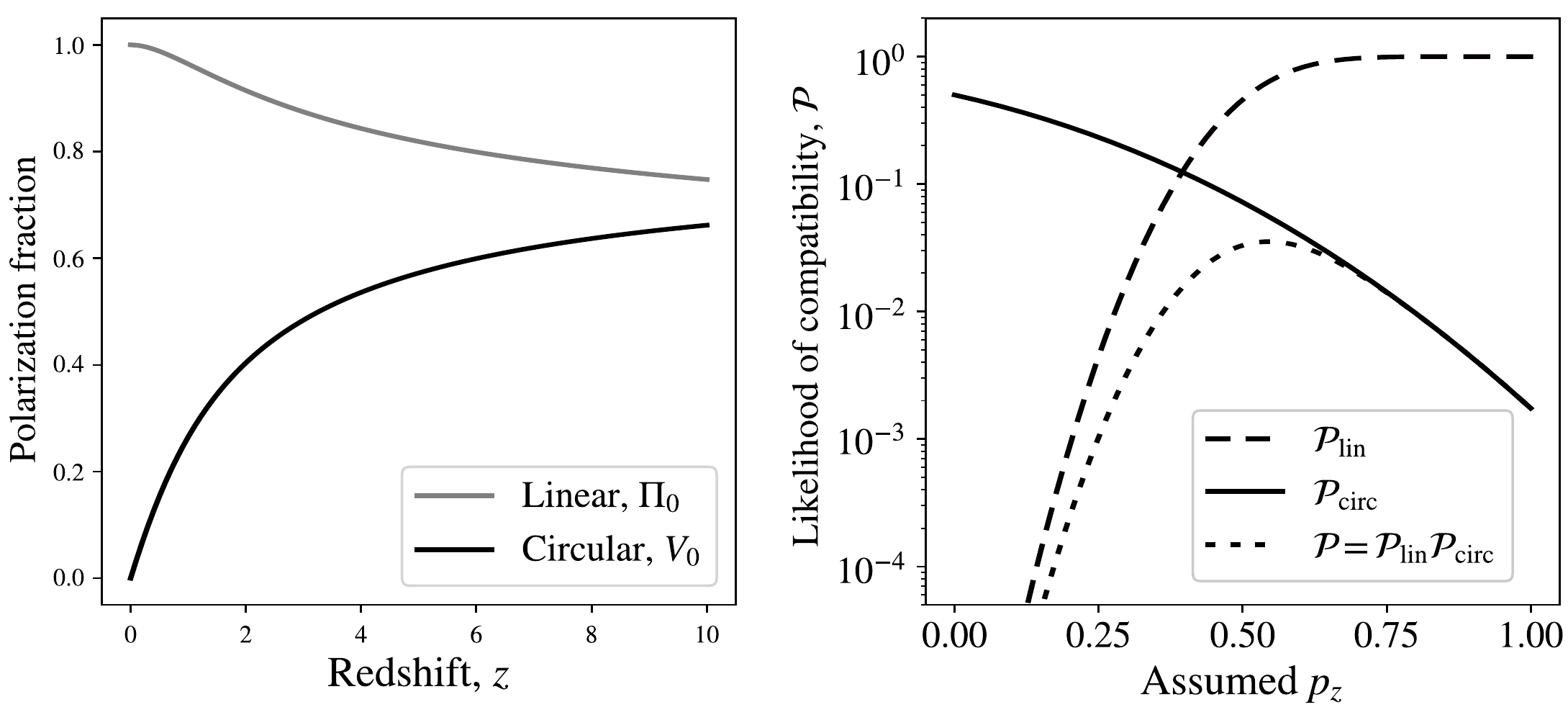}%
    \caption{\emph{Left:} Observable linear polarization decreases as photons propagate over a long distance while observable circular polarization increases.
    \emph{Right:} Probability to measure a given linear polarization fraction $\Pi = 0.5\pm 0.1$ and at the same time an upper limit $V \leq 0.01$ for a sample source, versus linear polarization at the source, $p_z$.
    Adapted from Gerasimov \etal\ (2021).\cite{gerasimov_etal_2021}
    }
    \label{fig:maximization}
\end{figure}

All analyses discussed so far, conservatively assume a high linear polarization at the source which would allow a high linear polarization to be observed even accounting for some level of LIV-induced depolarization.
For a relatively small number of AGN, upper limits of the circular polarization of optical light exist.
Most astrophysical emission processes are not expected to result in significant circular polarization.
Since circular polarization arises naturally from linear polarization in even-$d$ models (see Fig.~\ref{fig:maximization}), the absence of circular polarization combined with significant linear polarization provides a strong additional constraint.
By expanding the MCMC likelihood walk to include the circular polarization information, we were able to improve our $d=4$ SME constraints by an order of magnitude.\cite{gerasimov_etal_2021}

\section{Summary and Outlook}
Combining data from measurements of linear and circular polarization of optical light from distant astrophysical sources results in some of the strongest constraints on $d=4,5,6$ photon sector SME coefficients.
NASA's recently launched Imaging X-ray Polarimetry Explorer (IXPE) will provide significant $2-8\,\mathrm{keV}$ X-ray polarization measurements of several AGN with small systematic uncertainties.
Furthermore, the Compton Spectrometer and Imager (COSI) is currently under development, and once launched can provide polarization data at around $1\,\mathrm{MeV}$.
These high-energy polarization measurements will be even more sensitive Lorentz violating effects.

\end{document}